\documentclass[twocolumn,prb,showpacs]{revtex4}
\usepackage{graphicx}
\usepackage{dcolumn}
\usepackage{bm}
\begin{document}

\title{Resonant Light Absorption  by Semiconductor Quantum Dots  }
\author{I. G. Lang, L. I. Korovin}
\address{A. F. Ioffe Physical-Technical Institute, Russian
Academy of Sciences, 194021 St. Petersburg, Russia}
\author{S. T. Pavlov\dag\ddag}
\address{\dag Facultad de Fisica de la UAZ, Apartado Postal C-580,
98060 Zacatecas, Zac., Mexico;}
\address{\ddag
P.N. Lebedev Physical Institute, Russian Academy of Sciences,
119991 Moscow, Russia; pavlov@sci.lebedev.ru}

\begin{abstract}
The cross section of light absorption by semiconductor quantum dots is
calculated in the resonance with $ \Gamma_6 \times \Gamma_7 $ excitons in
cubic crystals $T_d $.  The interference of stimulating and induced electric
and magnetic fields is taken into account. The cross section is proportional
to the exciton nonradiative damping $ \gamma $ .
\end{abstract}


\maketitle

\section{Introduction}
When the size-quantized semiconductor objects (quantum wells,
quantum wires, quantum dots) are irradiated by light, elastic
light scattering and absorption intensify resonantly if the light
frequency $ \omega_l $ equals to the exciton frequency
 $ \omega_0 $. The resonant peak width is determined
by the exciton broadening $ \Gamma $, which consists of
nonradiative and radiative broadening, i. e., $ \Gamma =\gamma
+\gamma_r $. Therefore, measurements of light scattering and
absorption may serve as a convenient method of research of exciton
properties in the mentioned above objects. An important role of
the radiative damping $ \gamma_r $ was proved for the first time
in \cite{bb1,bb2,bb3}. Light reflection by some structures,
consisting of quantum wells, wires and dots was considered in
\cite{bb4}.

In the present work, a semiclassical method of retarded potentials
is applied for calculation of the light absorption cross section
of a quantum dot. The method allows to avoid using of boundary
conditions for electric and magnetic fields. That is especially
important in a case of quantum dots of arbitrary forms.

The method of the retarded potentials is described and applied to
the light scattering by a semiconductor quantum dot \cite{bb5}.
The calculations are performed for the resonance of stimulating
light with the exciton $ \Gamma_6 \times \Gamma_7 $ in cubic
crystals $T_d $. An exciton is formed by an electron  from the
twice degenerated conduction band $ \Gamma_6 $ and by a hole from
the twice degenerated valence band $ \Gamma_7 $, chipped off by
the spin - orbital interaction. The same excitons  are considered
in \cite{bb6}. The resonant light absorption is calculated below.

\section{Induced electric and magnetic fields}
Electron and hole wave functions  have a structure (we use
designations of \cite{bb7})
\begin{equation}
\label{1} \Psi_{c1}=iS\uparrow,~~~   \Psi_{c2}=iS\downarrow,
\end{equation}
\begin{eqnarray}
\label{2}
\Psi_{h1}={1\over\sqrt{3}}(X-iY)\uparrow-{1\over\sqrt{3}}Z\downarrow,\nonumber\\
\Psi_{h2}={1\over\sqrt{3}}(X+iY)\uparrow+{1\over\sqrt{3}}Z\downarrow.
\end{eqnarray}
Combining (1) and (2) in pairs, we obtain a four times degenerated
excitonic state, for which interband matrix elements of the
quasi-momentum operator are equal
\begin{eqnarray}
\label{3} {\bf p}_{cv1}&=&{p_{cv}\over\sqrt{3}}({\bf e}_x-i{\bf
e}_y),\nonumber\\
{\bf p}_{cv2}&=&{p_{cv}\over\sqrt{3}}({\bf e}_x+i{\bf
e}_y),\nonumber\\
{\bf p}_{cv3}&=&{p_{cv}\over\sqrt{3}}{\bf e}_z,\nonumber\\
{\bf p}_{cv4}&=&-{p_{cv}\over\sqrt{3}}{\bf e}_z,
\end{eqnarray}
where
\begin{equation}
\label{4} p_{cv}=i\langle S|{\hat p}_x|X\rangle,
\end{equation}
$ {\bf e} _x, {\bf e} _y $ and $ {\bf e} _z $ are the unit vectors
aligned parallel to the crystallographic axes.

An exciton $ \Gamma_6 \times \Gamma_7 $ is the most simple object,
unlike the excitons containing light or heavy holes. We obtain
that all the measurable values do not depend on the direction of
the crystallographic axes vectors, i. e., in the case of $
\Gamma_6 \times \Gamma_7 $ excitons, a crystal plays a role
 of an isotropic medium.

When the light irradiation is monochromatic, the stimulating
electric and magnetic fields can be written as
\begin{eqnarray}
\label{5} {\bf E}_0^{\pm}({\bf r}, t)&=&E_0{\bf
e}_\ell^{\pm}e^{i({\bf
k}_\ell{\bf r}-\omega_\ell t)}+ c.c.,\nonumber\\
{\bf H}_0^{\pm}({\bf r}, t)&=&E_0\nu [{\bf e}_z\times{\bf
e}_\ell^{\pm}]e^{i({\bf k}_\ell{\bf r}-i\omega_\ell t)}+ c.c..
\end{eqnarray}
The axis $z $ is aligned parallel to the light wave vector $ {\bf
k} _\ell $, whose magnitude equals $k_\ell =\omega_\ell\nu/c $, $
\nu $ is the light refraction coefficient, identical inside and
outside of the quantum dot. We use a circular polarization
\begin{equation}
\label{6}{\bf e}_\ell^{\pm}={1\over\sqrt{2}} ({\bf e}_x\pm i{\bf
e}_y).
\end{equation}
Induced electric and magnetic fields on the large distances from
the quantum dot, which determine light scattering and absorption,
are calculated in \cite{bb5}.

The values
\begin{equation}
\label{7}P({\bf k})=\int d^3r e^{-i {\bf k}{\bf r}}F({\bf r})
\end{equation}
are essential in the theory ($F ({\bf r}) $ is the real exciton
wave function at $ {\bf r} _e = {\bf r} _h = {\bf r}, ~~ {\bf r}
_e ({\bf r} _h) $ is the electron (hole) radius - vector). The
"envelope" wave function $F ({\bf r}) $ is calculated in the
effective mass approximation.

The electric and magnetic fields are calculated precisely, when
 $P ({\bf k}) $ depends on the magnitude $k $ only, i. e.,
\begin{equation}
\label{8}P({\bf k})=P(k).
\end{equation}
The condition (8) is carried out, if the function $F ({\bf r}) $
is spherically symmetric or if the quantum dot sizes $R$ are much
less than the stimulating light wave length, and
\begin{equation}
\label{9}P({\bf k})\simeq P(0).
\end{equation}
For example, condition (8) is carried out in the case of a
spherical quantum dot, limited by an infinitely high rectangular
potential barrier. Then  the "envelope" function
$$ F ({\bf r}) =F (r) = {1\over 2\pi R} {\sin^2 (\pi r/R) \over
r^2} \Theta (R-r)$$
corresponds to the lowest exciton energy level $ \hbar\omega_0 $,
and
\begin{equation}
\label{10}P(k)={2\over kR}\int_0^\pi dx
\sin{kRx\over\pi}\,{\sin^2x\over x},~~~P(0)=1.
\end{equation}
Under exact calculation of the induced fields we mean the account
of all the orders on the light - electron interaction (containing
the parameter $e^2/\hbar c $), what corresponds to the account of
all the processes of light reabsorption and reradiation. In
\cite{bb5}, the results
\begin{eqnarray}
\label{11} \Delta{\bf E}_{r\rightarrow\infty}({\bf r},
t)&=&-{3\over 4}E_0{\gamma_r\over k_\ell r}[({\bf e}_\ell{\bf
e}_s^{-}){\bf e}_s^{+}+({\bf e}_\ell{\bf e}_s^{+}){\bf
e}_s^{-})]\nonumber\\
&\times&{e^{i(k_\ell r-\omega_\ell t)}\over \omega_\ell-{\tilde \omega}_0+i\Gamma/2}+ c.c.,\nonumber\\
\end{eqnarray}
\begin{eqnarray}
\label{12} \Delta{\bf H}_{r\rightarrow\infty}({\bf r}, t)&=&{3i
\nu\over 4}E_0{\gamma_r\over k_\ell r}[({\bf e}_\ell{\bf
e}_s^{-}){\bf e}_s^{+}-({\bf e}_\ell{\bf e}_s^{+}){\bf
e}_s^{-})]\nonumber\\
&\times&{e^{i(k_\ell r-\omega_\ell t)}\over \omega_\ell-{\tilde \omega}_0+i\Gamma/2}+ c.c.,\nonumber\\
\end{eqnarray}
are obtained, $ {\tilde \omega} _0 =\omega_0 +\Delta\omega_0 $ is
the exciton energy renormalized by the electron-light interaction,
$ {\bf e} _s ^ {\pm} $ is the circular polarization vector, and
\begin{equation}
\label{13}\gamma_r={8\nu\over 9}{e^2\over\hbar
c}\left({p_{cv}\over
m_0c}\right)^2{\omega_\ell^2\over\omega_g}|P(k_\ell)|^2,
\end{equation}
$\hbar\omega_g $ is the band gap.

\section{The Pointing vector. The light scattering cross section}

On the large distances from the quantum dot, the Pointing vector
is equal
\begin{equation}
\label{14}{\bf S}_{r\rightarrow\infty}={\bf S}_0+{\bf
S}_{interf}+{\bf S}_{scat},
\end{equation}
where
\begin{equation}
\label{15}{\bf S}_0={c\over 4\pi}{\bf E}_0\times{\bf
H}_0={c\nu\over 2\pi}E_0^2{\bf e}_z,
\end{equation}
\begin{equation}
\label{16}{\bf S}_{interf}={c\over 4\pi}({\bf E}_0\times\Delta{\bf
H}+\Delta{\bf E}\times{\bf H}_0),
\end{equation}
\begin{equation}
\label{17}{\bf S}_{scat}={c\over 4\pi}\Delta{\bf
E}\times\Delta{\bf H}.
\end{equation}
With the help of (11) and (12), we obtain
\begin{equation}
\label{18}{\bf S}_{scat}={9\pi\over 4}S_0{\gamma_r^2\over(k_\ell
r)^2}{{\bf r}\over r}{|{\bf e}_\ell{\bf e}_s^{-}|^2+|{\bf
e}_\ell{\bf e}_s^{+}|^2\over (\omega_\ell-{\tilde
\omega}_0)^2+\Gamma^2/4},
\end{equation}
and the magnitude of the total flux of scattered light equals
\begin{equation}
\label{19}\Pi_{scat}={3\pi\over 2}S_0{\gamma_r^2\over
k_\ell^2}{1\over (\omega_\ell-{\tilde \omega}_0)^2+\Gamma^2/4}.
\end{equation}
In our calculations, we used the ratios
\begin{eqnarray}
\label{20}|{\bf e}_\ell^+{\bf e}_s^{-}|^2=|{\bf e}_\ell^-{\bf
e}_s^{+}|^2={1\over 4}(1+\cos\Theta)^2,\nonumber\\
|{\bf e}_\ell^+{\bf e}_s^{+}|^2=|{\bf e}_\ell^-{\bf
e}_s^{-}|^2={1\over 4}(1-\cos\Theta)^2,
\end{eqnarray}
where $ \Theta $ is the scattering angle, i. e., the angle between
the vectors $ {\bf k} _\ell $ and $ {\bf r} $. Then, the
integration on angles  $ \Theta $ is executed. Dividing $
\Pi_{scat} $ on the flux density $S_0 $ of stimulating light
energy and using the ratio $k_\ell=2\pi/\lambda_\ell $, we obtain
the total scattering cross section
\begin{equation}
\label{21}\sigma_{scat}={3\over
2\pi}\lambda_\ell^2{\gamma_r^2/4\over (\omega_\ell-{\tilde
\omega}_0)^2+\Gamma^2/4}.
\end{equation}
It follows from (21) that under condition $ \gamma\ll\gamma_r $ in
the resonance the total scattering cross section equals $ (3/2\pi)
\lambda_\ell^2 $, where $ \lambda_\ell $ is the light wave length.
Otherwise, at $ \gamma\gg\gamma_r $ the cross section in the
resonance decreases in $ (\gamma/\gamma_r) ^2 $ times. The damping
$ \gamma_r $ is determined in (13), and under condition
$R\ll\lambda_\ell $, the approximation  $P (k_\ell) \simeq P (0) $
is applicable.

\section{A role of the excitonic nonradiative  damping in light absorption}

Results for cross sections of light scattering by a quantum dot
are obtained with the help of the quasi-classical method and
coincide with results of the quantum perturbation theory
\cite{bb8} in the lowest approximation on the electron-light
interaction. However, the semiclassical method, consisting of the
calculation of electric and magnetic fields, allows to determine
also the light absorption section by quantum dots. The light
absorption is caused by nonradiative damping $ \gamma $ of
excitons and in frameworks of our problem is equal 0 at $ \gamma=0
$. Such result was obtained for a quantum dot in the case of
monochromatic irradiation \cite{bb2,bb9}. In the case of pulse
irradiation at $\gamma = 0 $ integral absorption is equal to zero
\cite{bb10,bb11,bb12}. The reason is that the energy dissipation
spent by light on the exciton creation is absent at $ \gamma=0 $,
and the energy comes back at the exciton annihilation. The process
of reabsorption and reradiation proceeds infinitely.

At calculation of the light absorption coefficient of a quantum
well, it was found out that it is necessary to take into account
the interference of stimulating and induced electromagnetic fields
(see, for example, \cite{bb10}).

Let us show that the last statement is true for the quantum dot
also.

\section{The interference contribution into the energy flux}

Let us calculate  the interference  contribution into the Pointing
vector. Substituting (5), (11) and (12) in (16), we obtain
\begin{eqnarray}
\label{22} &&{\bf S}_{interf}={\bf S}_z+{\bf S}_\perp,\nonumber\\
&&{\bf S}_z=-{3\over 4}{\gamma_r\over k_\ell r}S_0{\bf e}_z|{\bf
e}_\ell^+{\bf e}_s^{-}|^2\nonumber\\
&&\times\left({e^{i({\bf k}_\ell{\bf r}-k_\ell r)}\over
\omega_\ell-{\tilde \omega}_0-i\Gamma/2}+c.c.\right),~~~~~~
\end{eqnarray}
\begin{eqnarray}
\label{23}&&{\bf S}_\perp^\pm={3\over 4}{\gamma_r\over k_\ell
 r}S_0\nonumber\\&&\times \left({\bf e}_\ell^\pm({\bf
e}_\ell^\mp{\bf e}_s^{\pm})({\bf e}_s^\mp{\bf e}_z){e^{i({\bf
k}_\ell{\bf r}-k_\ell r)}\over \omega_\ell-{\tilde
\omega}_0-i\Gamma/2}+c.c.\right),
\end{eqnarray}
where the indexes $ + (-) $ correspond to the right (left)
polarization of the stimulating light.

Since expressions (22) and (23) correspond to the case
$r\rightarrow\infty $, it is obvious that only the angles $
\Theta\rightarrow 0 $ can contribute into the constant energy flux
because of the factor $e ^{i ({\bf k}_\ell{\bf r} -k_\ell r)} $.
However, in the RHS of (23) there is the factor $ {\bf e} _\ell
\mp {\bf e} _z $, which equals 0 at $ \Theta = 0 $. Therefore, the
constant energy flux on large distances from a quantum dot
corresponds only to the vector $ {\bf S} _ {z} $. Let us calculate
the energy flux
\begin{equation}
\label{24}\mathbf{\Pi}_{z}=\int ds {\bf S}_z,
\end{equation}
passing through a surface, perpendicular to the direction $z $ of
stimulating light in  time unit on a very large distance $z $
behind the quantum dot. The surface element is equal $ds =\rho
d\rho d\varphi $, and $ \rho=z\tan\Theta, r=z/\cos\Theta $. Having
executed integration on the angle $ \varphi $, which gives a
factor  $2\pi $, using (20) and passing on from the variable $
\rho $ to the variable $ \Theta $, we obtain
\begin{eqnarray}
\label{25}&&{\bf \Pi}_z=-{3\pi\over 8}{\gamma_rz\over k_\ell}{\bf
e}_z {S_0\over\omega_\ell-{\tilde
\omega}_0-i\Gamma/2}\int_0^{\pi/2}d\Theta\nonumber\\
&&\times {\sin\Theta(1+\cos\Theta)^2\over\cos^2\Theta} e^{ik_\ell
z[1-(1/\cos\Theta)]}+c.c..~~~~~~~~~
\end{eqnarray}
Further, let us pass to the variable $t =\cos ^ {-1} \Theta-1 $
from the variable $ \Theta $
\begin{eqnarray}
\label{26}&&{\bf \Pi}_z=-{3\pi\over 8}{\gamma_rz\over k_\ell}{\bf
e}_z {S_0\over\omega_\ell-{\tilde
\omega}_0-i\Gamma/2}\nonumber\\&&\times \int_0^\infty dt\Big
({2+t\over 1+t}\Big )^2 e^{-ik_\ell zt}+c.c..
\end{eqnarray}
At $z\rightarrow\infty $, having executed integration on $t $, we
obtain
\begin{equation}
\label{27}{\bf \Pi}_z=-{3\pi\over 2k_\ell^2}S_0{\bf e}_z
{\gamma_r\Gamma\over(\omega_\ell-{\tilde
\omega}_0)^2+(\Gamma/2)^2},
\end{equation}
independent on $z $. In (27), the contributions approaching to 0
at $z\rightarrow\infty $ are omitted, as well as the contributions
 oscillating very rapidly on $z $.

Thus, we obtain that the interference contribution to the Pointing
vector at $z\rightarrow\infty $ results into the constant energy
flux directed oppositely to the axis $z $. It means that the
energy flux of stimulating light decreases on the value $ {\bf
\Pi} _z $.

From expressions (25) and (26), it is obvious that at
$z\rightarrow\infty $ the basic contribution into the integral on
$ \Theta $ is brought in by the very small angles, when  $e^{i
({\bf k} _\ell {\bf r} -k_\ell r)} $ approaches to the unit.

The "transverse" component $ {\bf S} _ \perp $ of the interference
contribution to the Pointing vector  may be written down as
\begin{eqnarray}
\label{28} &&{\bf S}_{\perp}^\pm=-{3\over
8\sqrt{2}}S_0{\gamma_r\over k_\ell
r}(1+\cos\Theta)\sin\Theta\nonumber\\&&\times \left({e^{(i{\bf
k}_\ell{\bf r}-k_\ell r)}\over \omega_\ell-{\tilde
\omega}_0-i\Gamma/2}{\bf e}_\ell^\pm e^{\mp
i\varphi}+c.c.\right).~~~~~~~
\end{eqnarray}
In comparison with similar expression (22) for $ {\bf S} _ {z} $,
the RHS of (28) contains an additional factor $ \sin\Theta $,
which corresponds to the factor $ {\bf e} _z {\bf e} _s ^\mp =
(\pm i/\sqrt {2}) \sin\Theta $ from the RHS of (23). Due to this
additional factor, the contribution $ {\bf S} _ {\perp} ^ \pm $
approaches to zero in the area of very small angles $ \Theta $,
where $e ^ {i ({\bf k} _\ell {\bf r} -k_\ell r)} $ approaches to
the unit at $z\rightarrow\infty $. A direct calculation shows that
the flux approaches to zero at $z\rightarrow\infty $. Therefore,
it may be omitted.

\section{The light absorption cross section}

Since $ \Gamma =\gamma_r+\gamma $, the flux (27) consists of two
parts :
\begin{equation}
\label{29}{\bf \Pi}_z=-{\bf e}_z\Pi_{scat}-{\bf e}_z\Pi_{abs},
\end{equation}
where $ \Pi_{scat} $ is defined in (19),
\begin{equation}
\label{30}{\bf \Pi}_{abs}={3\pi\over 2}{S_0\over k_\ell^2}
{\gamma_r\gamma\over(\omega_\ell-{\tilde
\omega}_0)^2+(\Gamma/2)^2}.
\end{equation}
It is obvious that the  energy flux $ - {\bf e} _z\Pi _ {scat} $
compensates the total flux of the scattered energy, and $ - {\bf
e} _z\Pi _ {abs} $  is the energy flux absorbed by a quantum dot
during the time unit. Having divided (30) on the density $S_0 $ of
the stimulating energy flux, we obtain that the total light
absorption cross section results in
\begin{equation}
\label{31}\sigma_{abs}={3\over 2\pi}\lambda_\ell^2
{\gamma_r\gamma/4\over(\omega_\ell-{\tilde
\omega}_0)^2+(\Gamma/2)^2}.
\end{equation}
Comparing (31) with the expression (21) for the total scattering
section, we find that in the lowest approximation the scattering
section is of the second order on the electron-light interaction,
and the absorption section is of the  first order. In the
resonance, at $ \gamma\ll\gamma_r $ the ratio of the absorption
section to the scattering section equals $ \gamma/\gamma_r\ll 1 $.
Otherwise, at $ \gamma\gg\gamma_r $, the same ratio equals $
\gamma/\gamma_r\gg 1 $. The largest result for the absorption
section turns out at comparable values $ \gamma $ and $ \gamma_r
$. Thus, at $ \gamma =\gamma_r $ in the resonance,
\begin{equation}
\label{32}\sigma_{scat}^{res}=\sigma_{abs}^{res}={3\over
8\pi}\lambda_\ell^2.
\end{equation}
\begin{figure}
 \includegraphics [] {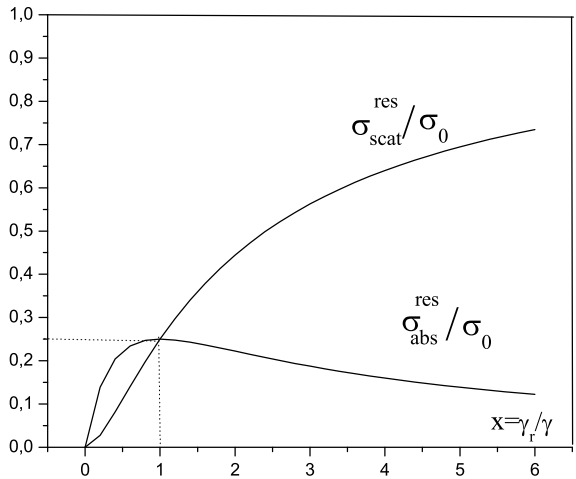} 
 \caption[*]{\label{Fig.eps}The cross sections $ \sigma_{scat}^{res} $ and  $
\sigma_{abs}^{res} $ as functions of the parameter $
\gamma_r/\gamma $. $\sigma_{0}=(3/2\pi)\lambda_{l}^{2}$.}
 \end{figure}

In Fig.1, the dependencies of $ \sigma_{scat}^{res} $ and  $
\sigma_{abs}^{res} $ on the parameter $ \gamma_{r}/\gamma $ are
represented.

In the resonance with the exciton $ \Gamma_6 \times \Gamma_7 $ in
any quantum dot, the result (31) for the absorption section is
true under condition $R\ll\lambda_\ell $, as well as in the case
of the spherically symmetric envelope wave function $F ({\bf r})
=F (r) $ of the exciton. If $R\geq\lambda_\ell $, the dependence
of $ \gamma_r $ on $k_\ell R $ may be essential (see (13) and
(10)).

\section{Conclusion}
Thus, the semiclassical method of the retarded potentials has
allowed to calculate induced electric and magnetic fields arising
at a light irradiation of a quantum dot if the stimulating light
frequency  is in the resonance with the exciton frequency. The
method has allowed to avoid using of boundary conditions for
fields. That has considerably facilitated the solution of the
problem. On the large distances from the quantum dot, fields  are
calculated precisely, i. e., without an expansion in series on the
electron-light interaction.

For excitons $ \Gamma_6 \times \Gamma_7 $ in cubic crystals of
$T_d $ class, the expressions  for light scattering and absorption
sections are obtained. It is shown that the account of the
interference contributions into the Pointing vector is necessary
for calculation of the light absorption

\end{document}